# Beyond the average: Detecting global singular nodes from local features in complex networks

L. da F. Costa[1(a)], F. A. Rodrigues[1], C. C. Hilgetag[2,3] and M. Kaiser[4,5,6]

[1] *Instituto de Física de São Carlos, Universidade de São Paulo - São Carlos, SP, PO Box 369, 13560-970 Brazil*
[2] *Jacobs University Bremen, School of Engineering and Science - Campus Ring 6, 28759 Bremen, Germany, EU*
[3] *Boston University - Sargent College, Department of Health Sciences - 635 Commonwealth Ave, Boston, MA 02215, USA*
[4] *School of Computing Science, Newcastle University - Claremont Tower, Newcastle upon Tyne, NE1 7RU, UK, EU*
[5] *Institute of Neuroscience, Newcastle University - Framlington Place, Newcastle upon Tyne, NE2 4HH, UK, EU*
[6] *Department of Brain and Cognitive Sciences, Seoul National University, College of Natural Sciences Shilim, Gwanak, Seoul 151-747, Korea*

**Abstract** – Deviations from the average can provide valuable insights about the organization of natural systems. The present article extends this important principle to the systematic identification and analysis of singular motifs in complex networks. Six measurements quantifying different and complementary features of the connectivity around each node of a network were calculated, and multivariate statistical methods applied to identify singular nodes. The potential of the presented concepts and methodology was illustrated with respect to different types of complex real-world networks, namely the US air transportation network, the protein-protein interactions of the yeast *Saccharomyces cerevisiae* and the Roget thesaurus networks. The obtained singular motifs possessed unique functional roles in the networks. Three classic theoretical network models were also investigated, with the Barabási-Albert model resulting in singular motifs corresponding to hubs, confirming the potential of the approach. Interestingly, the number of different types of singular node motifs as well as the number of their instances were found to be considerably higher in the real-world networks than in any of the benchmark networks.

While uniformity and regularity are important properties of patterns in nature and science, it is the minority deviations in such patterns which are often particularly informative. A prototypical example is the great importance given by animal perception to variations in signals, in detriment of constant stimuli. For instance, the outlines of shapes or objects play a much more important role in visual perception than uniform regions (see, for instance [1]). Similarly, our focus of visual attention is frequently driven by abrupt cues at the visual periphery (*e.g.* a dot of contrasting color, a small object movement or flashes). Even during saccadic eye movements (*i.e.*, abrupt, ballistic gaze displacements), small changes in the scene can be perceived [2].

There are many examples of the importance of minority deviations in other scientific areas, including mathematics (the importance of extremal values) and physics (*e.g.* singularities). In complex networks (*e.g.* [3,4]), the uniformity of connections is typically expressed with respect to the number of connections of each node, the so-called degree. Amongst the most uniformly connected types of networks are the random networks —also called Erdős-Rényi (ER) networks [5], characterized by constant probability of a connection between any pair of nodes. Because of its uniformity, the connectivity of this type of network can be well approximated in terms of the average and standard deviation of their node degrees, which is a consequence of its concentrated, Gaussian-like, degree distribution. Despite being understood in depth since the first half of the 20th century, ER networks play a relatively minor role as a model of natural phenomena, because it

(a)E-mail: luciano@if.sc.usp.br

is difficult to find natural systems that can be properly represented by the Poisson-based ER networks.

While global deviation from uniformity was ultimately the reason behind the success of complex network studies (*e.g.* [3]), a good deal of attention has focused on identifying uniformities in complex networks, such as node degree distributions (*e.g.* [6]) and simpler regions [7]. Only relatively few studies have targeted singularity identification in terms of local structure. For instance, Milo *et al.* [8] addressed the detection of motifs significantly deviating from those in random networks (see also [9]), while Travençolo *et al.* showed that the nodes with smallest outward accessibility tend to be at the border of networks [10].

Singular nodes can be understood as the most non-regular nodes in the networks. Because of their unique topological features, they are likely to play special roles in the networks. Hubs are a typical example of singular nodes present in many real-world networks [11]. It is interesting to note that, despite the ubiquity of hubs in network theory, no formal definition of these entities exist. If we consider hubs just as the nodes with degree larger than the majority of nodes in a given network, even random networks generated by the ER model may contain hubs as a consequence of random fluctuations. A possible definition of hubs can be obtained by considering them as the nodes that are singular in terms of node degree. In this way, the methodology that is presented in this article can be applied to determine objectively hubs in complex networks. More importantly, as we take into account additional measurements, other types of more general singular node motifs can also be identified. Indeed, we found eight types of individual node motifs —indicative of the relative position of nodes within the network, rather than of local multiple-node patterns, such as regular network motifs [8]. These singular node motifs were found to be characteristic for the analyzed networks.

The methodology proposed in the current article includes two main steps: i) several measurements [4] of the local connectivity are obtained for each node; then ii) singular motifs detection methodologies from multivariate statistics and pattern recognition (*e.g.* [12]) are applied to identify the nodes exhibiting the greatest local structure deviations from the whole set of network vertices. The adopted measurements include a) the normalized average degree, $r(i)$, b) the coefficient of variation of the degrees of the immediate neighbors of a node, $cv(i)$, c) the clustering coefficient, $cc(i)$ [13], d) the locality index, $loc(i)$, which is an extension of the matching index (*e.g.* [14]) and takes into account all the immediate neighbors of each node, instead of individual edges; e) the hierarchical clustering coefficient of level two, $cc_2(i)$ (*e.g.* [15]); and f) the normalized node degree, $K(i)$. Each node is therefore mapped into 6-dimensional vectors $\vec{X}$, which "live" in the 6-dimensional feature space, defining distributions of points.

A number of concepts and methods have been developed that allow the identification of singular motifs in data sets (*e.g.* [12]). The methodology proposed and used in the current article to obtain the singular motifs involves the following three steps: i) projection of the six-dimensional vertices feature vectors into two-dimensional (2D) space, using principal component analysis [16] to remove correlations between measurements; ii) identification of singular nodes in the 2D space by considering the Parzen windows method for estimating the probability densities (the singular nodes correspond to the vertices leading to the smallest probability densities); and iii) supervised classification of the singular nodes into categories, according to proximity in $M$-dimensional space.

The first step, the projection of the feature vector of each vertex onto the 2D plane, can be obtained by principal component analysis (*e.g.* [4,12,17]). This method transforms the data into a new coordinate system (through a rotation) such that the greatest variances are concentrated along the first coordinate axes. To perform this method, the covariance matrix $\Sigma$ of the data is estimated and the eigenvectors corresponding to the largest absolute eigenvalues are calculated and used to define a linear transformation projecting the cloud of points into a space of reduced dimensionality (2D in the current case).

The identification of the singularities purely by visual inspection can produce inaccurate results because of the form of the distribution of points in the projection. In the present work, the singular nodes are determined in a quantitative fashion by considering the probability density in 2D space as estimated by the non-parametric Parzen windows approach [18,19]. This method involves adding a Gaussian function at each data point in 2D space in order to allow the interpolation of the probability density. The considered standard deviation of the Gaussian is equal to $\sigma_{xx} = \sigma_{yy} = 0.05$. In this way, points that are isolated from other data points will have small values of probability as there is no overlap in probabilities. Recall that the proximity between the points in the 2D space (implying higher spatial density) reflects the similarity of topological features of the respective nodes (*e.g.* [7]). In this work we consider as singular nodes the $w = 20$ vertices with the smallest probabilities.

Having identified the singular nodes, they are organized into categories sharing structural properties in the network. This grouping was done by considering $k$-means, which is a clustering method used to group objects with similar features [16]. The clustering is illustrated in fig. 1, at the 2D scatterplot. In order to be able to define a more complete description of such categories, it is necessary to return to the $M$-dimensional original space of measurements (here: $M = 6$). The objective is to obtain classification regions in this space so that unclassified nodes (from the reference or other networks) can be assigned to the singular motif categories. Because most network nodes are not singular, it is important that the singular motif categories in the $M$-dimensional space do not cover that whole space. In other words, the partitioning of the $M$-dimensional space involves the identified singular node motif categories plus the "non-singular" categories.

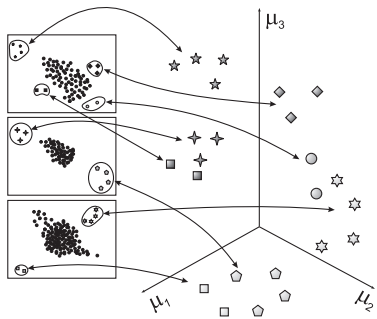

Fig. 1: Illustration of the proposed method to identify the singular motifs. First, the feature vectors of the vertices are projected onto the two-dimensional space by principal component analysis. Next, singular nodes are identified in this space by considering the probability distribution generated by the Parzen windows approach. These singular nodes are then grouped into categories by applying $k$-means clustering (each marked region on the plane). Finally, the classes of singular motifs are mapped back into the original $M$-dimensional measurement space and grouped according to the Mahalanobis distance.

This can be done by considering the original feature vectors of each singular node. Such a backmapping into the original measurement space defines clusters of singular motif categories which may exhibit overlap (recall that the grouping of the singular nodes was performed in the 2D space). In order to obtain non-overlapping singular motif regions, we perform a Voronoi expansion by considering the Mahalanobis distance, which is given by

$$D(X) = \sqrt{(\vec{X} - \vec{\mu})^T \Sigma^{-1} (\vec{X} - \vec{\mu})}, \qquad (1)$$

where $T$ stands for matrix transposition, $\vec{\mu}$ is the average of the feature vectors of each class and $\Sigma$ is the respective covariance matrix. The Mahalanobis distance [20], instead of the Euclidean distance, was adopted because the distributions of vertices in the original feature can be skewed and elongated. Recall that the Mahalanobis Voronoi expansion involves the average vector and covariance matrix of each singular motif category. In order to spatially constrain the singular motif regions, therefore accounting for the "non-singular" category (corresponding to the points in the $M$-dimensional space not covered by the Voronoi expansion), the Mahalanobis distances are constrained to values smaller than a threshold $d$. The categories which resulted in too close proximities (in the Mahalanobis sense) are grouped into a single region. In the end, this procedure partitions the original $M$-dimensional space into regions for the singular motif categories plus the "non-singular" region.

Figure 2 presents the projections of the original measurement spaces (obtained by principal component analysis) for (a) the US air transportation network, (b) the yeast protein-protein interaction network, and (c) for Roget's thesaurus network. To remove scaling biases, the six adopted measurements were standardized (*e.g.* [12]), which was accomplished by subtracting the average from each measurement and dividing by the standard deviation. Each of the two projection axes corresponds to a linear combination of the six original measurements. The respective probability density estimations are shown in fig. 2. The singular nodes correspond to the 20 vertices with the smallest probabilities.

The identified singular nodes were grouped into classes according to their proximity in the projections. After mapping these categories back into the original 6-dimensional measurement space, those categories which were too close to one another were merged (see fig. 1), resulting in eight final singular motifs, as shown in fig. 3. Why were just eight different singular motifs found, out of the many ($2^6 = 64$) potential groups? Among several possible explanations we have that: a) some groups are absent (due to skewed feature distributions) b) some groups are present but not included in the top eight singularities, and c) some features strongly correlate with each other leading to the merger of potential singular motifs. For example, if a minimum feature A correlates with a maximum in feature B (negative correlation), singular nodes may form a joint group AB. However, if all features are statistically independent and distributions are non-skewed, all potential groups of singular nodes should also occur in the top list. In short, considering absent singular motifs can provide additional information about the nature of network connectivity, *i.e.* a network does not present singular motifs when its structure is highly uniform, as we will demonstrate next for the ER network.

Having obtained the partition of the original six-dimensional measurement space into the eight outlier categories plus the "non-outlier" class, it is possible to categorize all nodes in any given network into these nine categories. Figure 4 presents the distributions of the singular motifs in the three real networks and in three benchmark networks: the random graphs of Erdős and Rényi (ER) [5], the small-world graphs of Watts and Strogatz (WS) [13], and scale-free networks of Barabási and Albert (BA) [6]. We can see that the singular motifs are more abundant in real networks than in the networks generated by the benchmark models. Also, the number of singular motifs is much smaller than the number of vertices classified as "non-singular". The networks generated by the ER and WS models tended not to produce singular motifs because of the regularity of their structures. On the other hand, the network constructed by the BA model contains a significant number of global relay hubs. Interestingly, these singular motifs turned out to correspond precisely to hubs with small clustering coefficient. Therefore, the proposed methodology for outlier detection provided a quantitative and objective way to identify the hubs as special motifs in the case of the BA model. The US air transportation network presented local relay hubs, cluster-tails and tail; the protein-protein interaction network global relay hubs, local integrators and cluster-tails; while Roget's thesaurus network included integrator, community center, cluster-cluster and tail singular motifs.

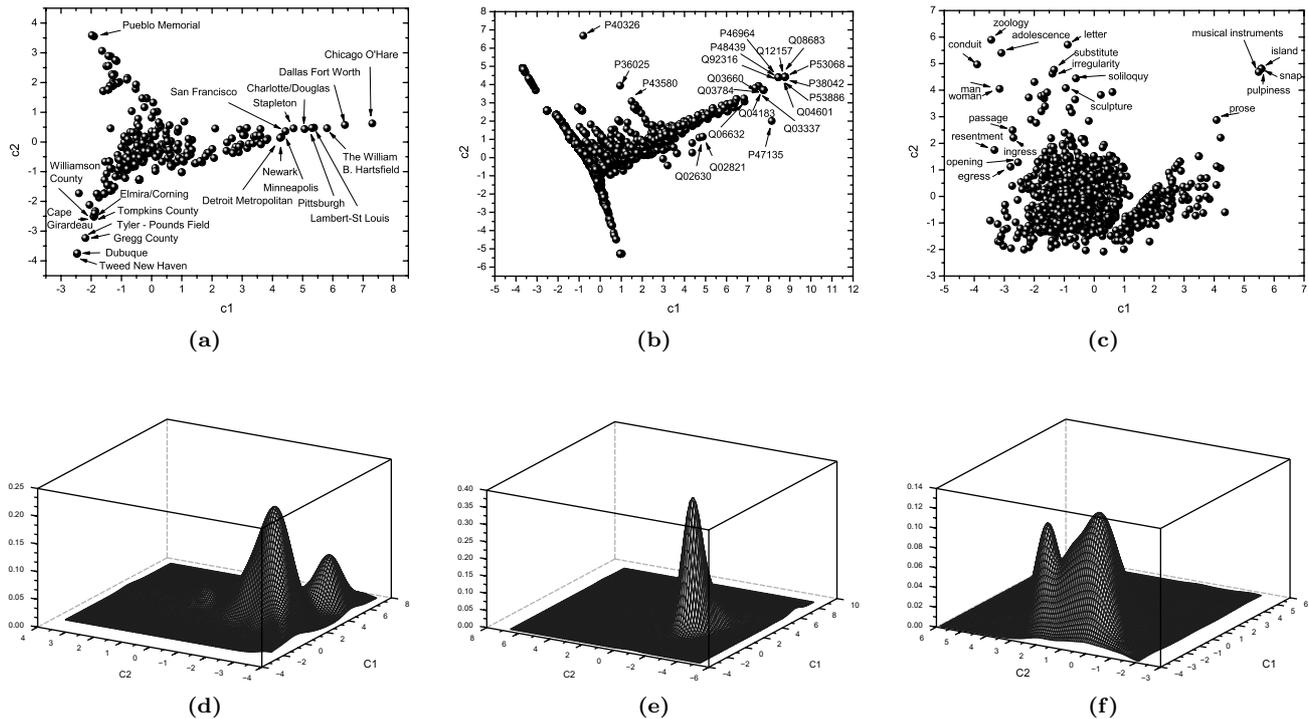

Fig. 2: The feature space obtained by principal component projections of the six-dimensional measurement vectors of the networks of (a) air transportation, (b) protein-protein interactions, and (c) Roget's thesaurus. The respective probability density functions are shown in (d), (e), and (f). The singular nodes are indicated by arrows.

Table 1 presents the 20 identified singular nodes in each real network. In the case of the US air transportation network, while all airports of class B (local relay hub) are international, the airports of class C (local integrator) are regional. The international airports can be understood as the hubs with values of the clustering coefficient larger than zero. This structure reflects the importance of such airports, since they concentrate high traffic. The airports of class F (cluster-tail) are characterized by large clustering coefficient and high variation in the degree of their neighbors, tending to be linked to a hub airport and an airport with few connections (see fig. 3, class F). For instance, the airport Dubuque Regional is connected to the Waterloo Municipal Airport and to Chicago O'Hare International Airport. The same occurs to the remainder type-F airports, but exceptionally, the Gregg County and the Tyler-Pounds Field airports connected to the Dallas Fort Worth International airport. The airport Pueblo Memoria, belonging to class H (tail), is a small airport mostly used for general aviation and served by just one commercial airline. This airport is linked to the Colorado Springs Airport, which is connected to seven airports that share links ($cc_2 = 1$). In this case we have a configuration where a small airport is connected to a larger regional airport. Therefore, the obtained singular motifs corresponded to particular functional types of airports.

In the case of protein-protein interactions, the 20 singular nodes fall into classes A (global relay hubs), C (local integrators) and F (tails). The global relay hubs present a small value of the clustering coefficient. According to previous studies [21], removal of hub proteins would tend to be lethal. This trend was only partially observed in our results. For instance, despite the fact that the protein P47135 (protein JSN1) is highly connected, it is considered viable. On the other hand, the protein Q02821 is lethal and promotes the docking of import substrates to the nuclear envelope [22]. The protein Q02630 has unknown viability. Among the 14 proteins considered local integrators, 11 are lethal (only the proteins P53068, P48439 and Q92316 are viable). Particularly, the proteins P38042, P53068, P53886, Q04601, Q08683 and Q12157 participate in the anaphase promoting complex/cyclosome which is involved in cell division [22]. These proteins are centers of cliques (set of fully connected subgraphs), presenting the same local structure with connectivity equal to 10 and clustering coefficient equal to 1. Similarly, the proteins Q03337, Q03660, Q03784, and Q04183 are components of the TRAPP II complex, which seems to play a role in intra-Golgi transport and in meiosis following DNA replication [22]. The proteins P46964, P48439, and Q92316 also have similar functions, being required for a maximal enzyme activity. The protein Q06632 is involved in coupling transcription termination and mRNA 3'-end formation. Therefore, the local integrators have similar functions. Despite the fact that such proteins are not hubs, they tend to be lethal, which suggests that the connectivity is not the only topological feature fundamentally associated to lethality of proteins [23]. The fact that the local integrator singular motifs have high clustering coefficient and high locality index indicates that such proteins can be found in the middle of communities. The cluster-tail proteins have small degree, high clustering coefficient, high normalized average neighboring degree and high coefficient of variation. All these proteins (P36025, P43580, and P40326) are viable and considered uncharacterized proteins [22]. They tend to be connected to a hub protein and to a less well connected protein. Therefore, these results suggest a relationship between the function and local singularity structure of proteins.

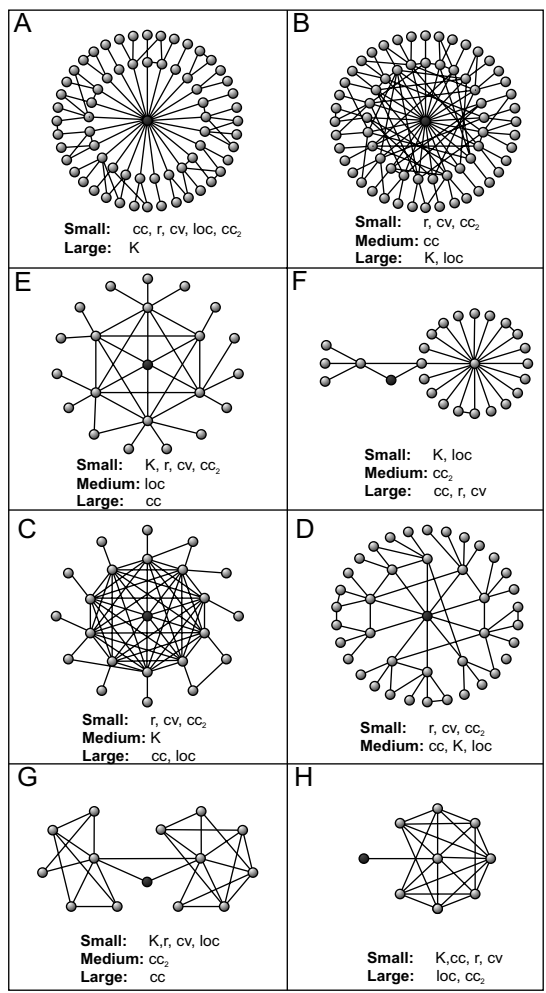

Fig. 3: New singular network motifs defined by the singular nodes identified in the three considered real-world networks. The singular node is the darkest node in each motif. We named these singular motifs (A) global relay hub, (B) local relay hub, (C) local integrator, (D) integrator, (E) community center, (F) cluster-tail, (G) cluster-cluster, and (H) tail.

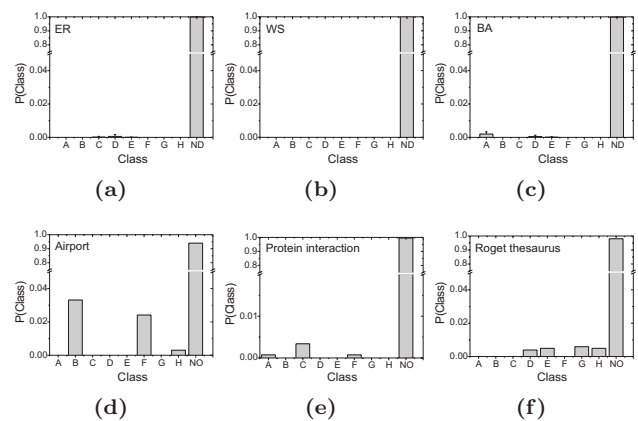

Fig. 4: The distribution of the singular motifs found in networks generated by (a) the Erdős-Rényi model, (b) the Watts-Strogatz model and (c) the Barabási and Albert model; as well as in real networks, namely (d) the US air transportation network, (e) the protein-protein interaction of the yeast *S. cerevisiae*, and (f) Roget's thesaurus network. The distributions for the models were obtained from 100 realization, where each network was composed of $N = 1,000$ nodes with average degree $\langle k \rangle = 4$. NO indicates non-singular nodes and each letter corresponds to the singular motifs presented in fig. 3.

The Roget's thesaurus network presents four types of singular motifs. All these classes are composed of sparsely connected words with specific local structures. The words of classes E (community center) and G (cluster-cluster) are similar, having small degree, but differing mainly in the average normalized neighboring degree. The community center words tend to be regular structures, presenting high clustering coefficient, small normalized average neighboring degree and a small coefficient of variation. Moreover, the words "man", "woman" and "adolescence" are interconnected belonging to the same community [24]. The words of class D (integrator) can be considered as internal to the communities, since their clustering coefficient is larger than zero. For instance, the outlier "resentment" is connected to words associated with hostility, such as "violence", "revenge" and "disrespect". The remaining words of class E (community center), "passage", "ingress", "egress" and "opening" are related to the same meaning. The cluster-cluster words are characterized by small degree, large clustering coefficient and hierarchical clustering coefficient of level two significantly larger than zero. In this case, the singular nodes of this class are connected with two words $i$ and $j$ of similar meaning, and the words connected to $i$ do not have semantic relation to the neighbors of $j$. For instance, the outlier "letter" is connected to the words "writing" and "printing". While the former is also connected to words associated with investigation, as "evidence" and "indication", the latter is associated with publishing, as "engraving" and "publication". Therefore, the cluster-cluster singular motifs tend to appear between communities. On the other hand, tails present degree one and a high clustering coefficient of level two. Thus, these singular words tend to be specific, being connected to one word whose neighbors do not have any association with that outlier word. For instance, the word

Table 1: The 20 most singular nodes obtained for the US air transportation network, the protein-protein interaction network of *S. cerevisiae* and the network of Roget's thesaurus, as well as respective node classifications.

| Airports | Class |
|---|---|
| Chicago O'Hare Intl., Dallas Fort Worth Intl., Pittsburgh Intl., Lambert-St Louis Intl., Charlotte/Douglas Intl., The William B. Hartsfield Intl., Stapleton Intl., Minneapolis-Saint Paul Intl., Detroit Metropolitan Wayne County, Newark Intl., San Francisco Intl. | B |
| Dubuque Reg., Tweed New Haven Reg., Gregg County, Tyler - Pounds Field Reg., Tompkins County Reg., Elmira/Corning Reg., Williamson County Reg., Cape Girardeau Reg. | F |
| Pueblo Memorial | H |

| Proteins | Class |
|---|---|
| P47135, Q02821, Q02630, P38042 P53068, P53886, Q04601, Q08683 | A |
| Q12157, P46964, P48439, Q92316, Q03337, Q03660, Q03784, Q04183, Q06632 | C |
| P36025, P43580, P40326 | F |

| Words | Class |
|---|---|
| resentment, passage, ingress, opening, egress | E |
| zoology, adolescence, man, woman | D |
| soliloquy, irregularity, substitute, letter, absence of intellect, sculpture | G |
| prose, musical instr., island, pulpiness, snap | H |

"prose" is connected to "poetry", which is connected to "melody", "music", "musician" and "voice". These neighbors of "poetry" have an association with music, but do not bear any semantic relation to it.

The extension of the current work to further measurements and different networks is straightforward and promising. Depending on the considered networks, other types of singular motifs may be obtained than the eight classes identified here. Moreover, the proposed method allows to classify the nodes according to local features (defined by the considered measurements) and to investigate the relation between the structure and function of complex networks.

∗ ∗ ∗

LdFC is grateful to CNPq (301303/06-1) and FAPESP (05/00587-5) for financial support. FAR is grateful to FAPESP (07/50633-9). MK was supported by the CARMEN e-science project (www.carmen.org.uk) funded by the EPSRC (EP/E002331/1).